\documentclass[%
 reprint,
 amsmath,amssymb,
prl,
]{revtex4-1}

\usepackage{graphicx}% Include figure files
\usepackage{dcolumn}% Align table columns on decimal point
\usepackage{bm}% bold math
\usepackage{color}
\usepackage{epstopdf}
\usepackage{gensymb}
\usepackage[english]{babel}
\usepackage{float}
\usepackage{xr}
\usepackage{sidecap}
\usepackage{mathrsfs,amsmath}
\usepackage[normalem]{ulem}

\begin{document}

\preprint{APS/123-QED}

\title{Observation of edge waves in a two-dimensional Su-Schrieffer-Heeger acoustic network}
% Force line breaks with \\
\author{Li-Yang Zheng}
%\email{liyang.zheng.etu@univ-lemans.fr}
\author{Vassos Achilleos}
%\email{florian.allein@univ-lemans.fr}
\author{Olivier Richoux}%
\author{Georgios Theocharis}
% \email{georgiostheocharis@gmail.com}
\author{Vincent Pagneux}
\affiliation{%
LAUM, UMR-CNRS 6613, Le Mans Universit\'e, Av. O. Messiaen, 72085 Le Mans, France}%

%\affiliation{
%\affiliation{
%\affiliation{%


\date{\today}% It is always \today, today,
\begin{abstract}
In this work, we experimentally report the acoustic realization the two-dimensional (2D) Su-Schrieffer-Heeger (SSH) model in a simple network of airchannels. We analytically study the  steady state dynamics of the system using a set of discrete equations for the acoustic pressure, leading to the 2D SSH Hamiltonian matrix without using tight binding approximation. 
%The topological edge waves supported by the 2D SSH model are obtained for the acoustic network both analytically and  numerically. 
By building an acoustic network operating in audible regime, we experimentally demonstrate the existence of topological band gap. More supremely, within this band gap we observe the associated edge waves even though the system is open to free space. Our results not only experimentally demonstrate topological edge waves in a zero Berry curvature system but also provide a flexible platform for the study of topological properties of sound waves. 

%The appearance of topological edge waves in the acoustic network 
\end{abstract}
\maketitle
The study of topological insulators has been attracting a lot of attention in %\sout{the last years}
recent years due to their appealing property for the control of wave propagation~\cite{HasanRMP, QiRMP}. Among other properties, topological insulators exhibit non-trivial topological phases, leading to the existence of robust edge states on the boundaries/interfaces~\cite{soljacic2014, Ma2019}. 
Various systems exhibiting non-trivial topological phases  have been investigated~\cite{RechtsmanNature,YvesNC,LuValleyVortex,DongValley,WangTRB,YangTA, WangTIPRL,Nash,KaneQSH,BHZ,KhanikaevTI,MousaviNC,WuPRLQSH, HuberSciences, HeNP,ZhengPRBRC}.
%Various types of schemes with non-trivial topological phases has been investigated in different systems []. 
Previous studies have shown that Chern insulators and $Z_2$ topological insulators possess a non-trivial topology phase stemming from non-vanishing Berry curvatures~\cite{WangTRB,YangTA,WangTIPRL,Nash,KaneQSH,BHZ,KhanikaevTI,MousaviNC,WuPRLQSH, HuberSciences, HeNP,ZhengPRBRC}. On the other hand, it has been recently reported that a topological phase can also appear in systems even in the absence of Berry curvature~\cite{KaneTBM,Xiao1DSSH}. 
This new interesting scheme has been found in the two-dimensional (2D) Su-Schrieffer-Heeger (SSH) model, which is a 2D extension of the 1D SSH chain, %\sout{where alternating strengths of bonds between identical atoms stand in both $x-$ and $y-$directions}
with alternating strengths of bonds connecting identical atoms in both $x-$ and $y-$directions~\cite{WakabayashiPRL,WakabayashiPRB,Xie2018,Liu2019}.

The topological phase in the 2D SSH model can be characterized by
2D Zak phase and topological edge states are consequently predicted on the boundaries of these structures. 
Due to the  difficulty in tuning the lattice couplings, at will in the quantum world, most of the attention has been devoted to study its analogues in classical systems, including photonics~\cite{WakabayashiPRB,Xie2018}, and electrical circuits~\cite{Liu2019}. %(phononic: WE THINK THERE IS NO PAPER ON THAT), 
However, acoustic analogues of the 2D SSH model have not been reported so far. 
The experimental realization of the 2D SSH model in acoustics not only can provide a simple and versatile platform for the study of topological edge waves, but also opens perspectives for activities involving other novel topological phase, such as high order topological insulators~\cite{Xie2018b,Chen2019,Ota2018,XueNM, ImhofNP, Marques}.

The analyses of topological insulators is usually performed
starting from a discrete %\sout{lattice }
model with special lattice symmetry. %\sout{ properties}. 
However, the majority of systems in different domains of physics are described by continuum models associated with partial differential equations. One of the most popular techniques to bridge the gap between the continuous and the discrete models is the Tight-Binding approximation (TBA)~\cite{ashcroft,lidorikis1998}. The original idea of TBA is to singularise discrete points in space at places where the continuous field is localised, thus it is naturally associated with resonating scatterers. This approximation technique can be rigorously applied by using Wannier functions basis, leading to the evaluation of delicate overlap integrals~\cite{lidorikis1998,marzari2012,pedersen1991}. 
In practice, for the application to topological insulators where the medium is periodic, it appears that the TBA provides generic 
discrete equations or dispersion relations with coupling coefficients that can be fitted with results from numerical simulation of the continuous problem~\cite{Ni2017}. 
An alternative approach which we utilize below, in the spirit of quantum graph theory~\cite{Kuchment2007}, is to directly obtain
a discrete model from a continuous system with coveted hopping coefficients, by combining wave propagation properties and geometrical characteristics of the system.
One advantage of the proposed %\sout{latter}
approach is that it is constructive in the sense that: the hopping coefficients can be prescribed at will, as opposed to the common practice in continuous systems where the hopping coefficients are~\textit{a posteriori} calculated (e.g. by fitting or using asymptotic methods)~\cite{Ni2017, craster2017,  Mei2012}.
%wheb a structure is already constructed.
% the desired discrete model 
%
%can be 
%exactly mapped at any prescribed frequency of the continuous system, allowing simple closed form expressions of the hopping coefficients \cite{craster2017}. 
%This is in contrast to t
%
%
%since it is asymptotic (with the slenderness as the small parameter), 
In addition, avoiding the use of resonating scatterers, the obtained discrete model can be tuned to be valid in a broadband range of frequencies.

In this work, we theoretically and experimentally study an acoustic 2D SSH network composed of simply connected air channels. Using the conservation of flux at the network junctions, we derive a set of discrete equations and subsequently map the acoustic system to the 2D SSH Hamiltonian. 
%We develop an analytical discrete model to describe the sound wave propagation in the acoustic network. 
%Then, a dynamical matrix equation for sound wave properties in the network is derived, which has the same form as the 2D SSH Hamiltonian.
The validity of our theoretical model
is checked by comparison of the dispersion relation with numerical simulations.  In addition, both our theoretical model and the numerical simulations  predict the appearance of topological edge states. Then
% We theoretically and numerically analyze the dispersion curves of the network, and we predict the appearance of topological edge waves on the boundaries of the network.
an experimental implementation of the 2D SSH is achieved in the audible regime and it contains two different boundaries, one supporting edge waves and the other not. The propagation of topological edge waves is experimentally observed, recovering the characteristic profile of the SSH edge modes and exhibiting localization only on the prescribed boundaries of the network.

\begin{figure}[t]
\includegraphics[width=8cm]{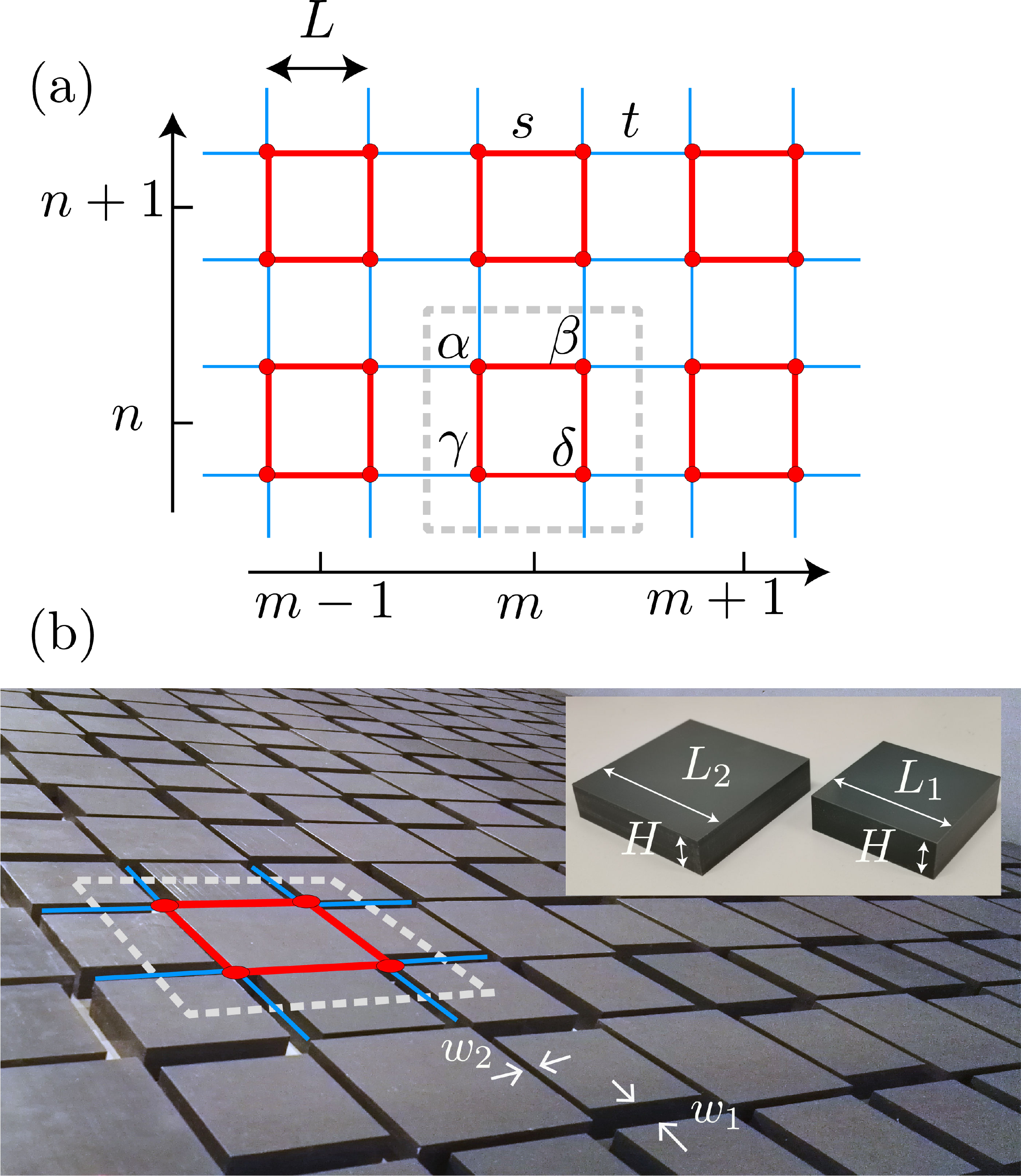}
\caption{\label{fig1} (a) Schematic presentation of the 2D SSH model, which is a square lattice of lattice constant $2L$. The unit cell containing four nodes $\alpha$, $\beta$, $\gamma$ and $\delta$ is highlighted by a gray-dashed box. The red and blue bonds represent the intracellular and intercellular couplings.
(b) A glance of the 2D acoustic network, which is a realization of the 2D SSH model for sound waves. The insert shows the two building blocks of the acoustic network. In the network, the intracelluar and intercelluar couplings can be achieved by changing the width of the air channels $w_2$ and $w_1$, respectively.}
\end{figure}

The 2D SSH model is depicted in Fig.~\ref{fig1}(a), where identical nodes are arranged in a square lattice with lattice constant $2L$. % in both $x-$ and $y-$directions. 
The unit cell containing four nodes (marked as $\alpha$, $\beta$, $\gamma$ and $\delta$) is indicated by a gray-dashed box in Fig.~\ref{fig1}(a). The intracellular (intercellular) hoppings, i.e., couplings of nodes within (between) unit cells, are denoted by $s$ ($t$) as the red (blue) bonds in Fig.~\ref{fig1}(a).
%Previous works have reported that the 2D SSH model exhibits a topological nontrivial phase characterized by the wave polarization described by an extended Zak phase in 2D. 
The acoustic realization of the 2D SSH model consists of a network structure which is shown in Fig.~\ref{fig1}(b). As it can be seen, the network is composed by two types of rigid square blocks with widths $L_1$, $L_2$, and height $H$ as shown in the inset of Fig.~\ref{fig1}(b). By placing the two blocks centered in a square lattice substrate with a lattice constant $2L$, two types of airborne channels of width $w_1$ and $w_2$ are fabricated. Then, covering the top of the structure with an additional plate, these channels form rigidly closed waveguides for the acoustic waves and the 2D SSH acoustic network is constructed. A single unit cell of the network is marked by a gray-dashed box in Fig.~\ref{fig1}(b), and the four junctions between the air-channels correspond to the nodes of the unit cell in Fig.~\ref{fig1}(a). 
The sound pressures at the each junction is coupled with those on its neighboring junctions through the channels of  alternating widths $w_1$ and $w_2$. Thus, this coupling between neighboring junctions can be easily tuned solely by the widths  $w_{1,2}$  which, as we will show below, play the same role as the intercellular and intracellular hoppings in the 2D SSH model, which we call $s$ and  $t$ respectively.

\begin{figure}[t]
\includegraphics[width=8cm]{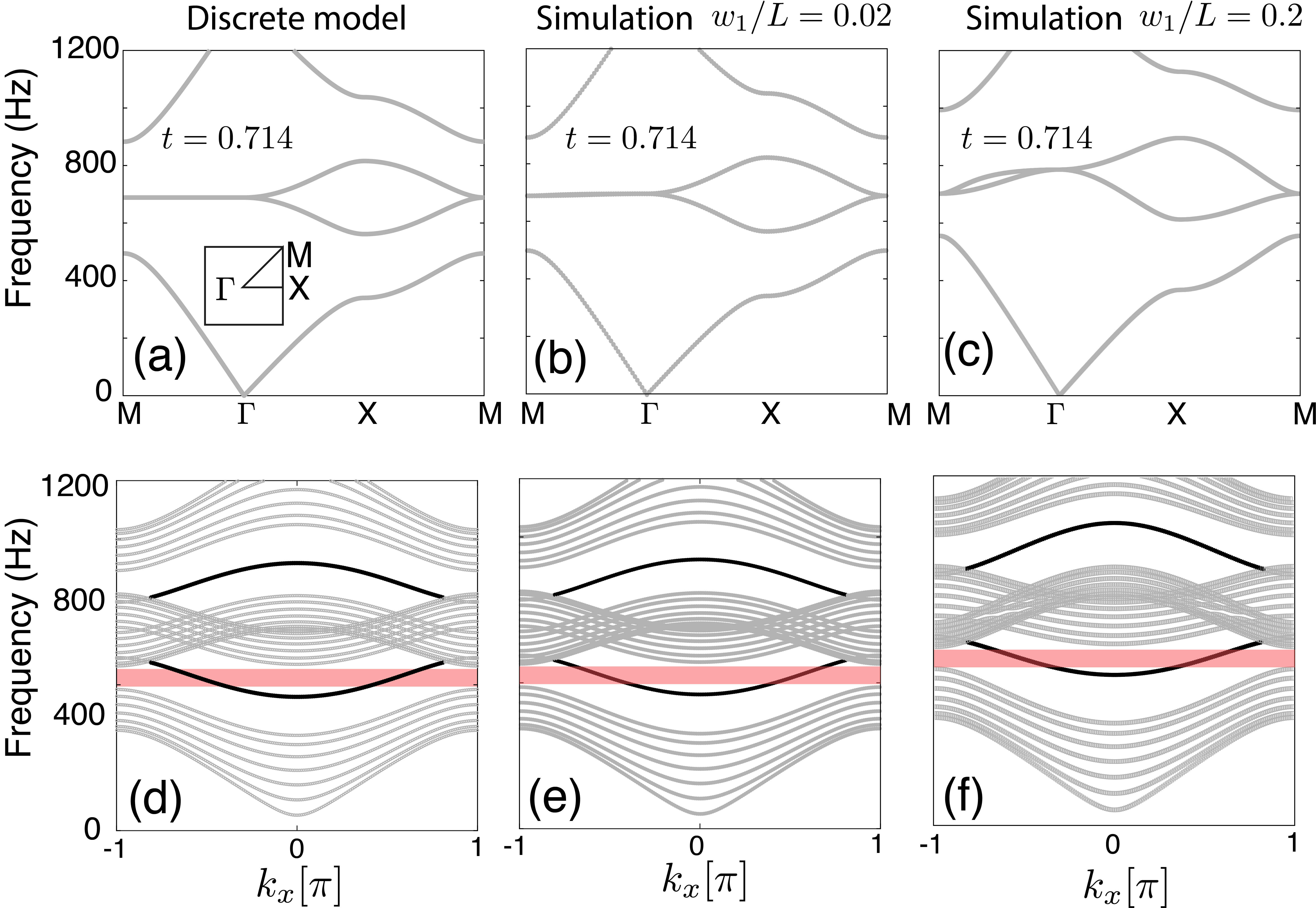}
\caption{\label{fig2} The dispersion relations of the SSH network when the intercelluar coupling $t=0.714$. (a)-(c) show the results for bulk modes, and (d)-(f) show the results for edge modes by considering a supercell consisting of $8$ unit cells. Lines in gray represent the bulk modes, and lines in black correspond to the edge modes. (a), (d) Theoretical results obtained from the network model. The insert in (a) presents the first Brillouin zone. (b), (e) Numerical results by considering the air channels of size $w_1/L=0.02$. (c), (f) Numerical results by considering the air channels of size $w_1/L=0.2$. } % (e) Considered interactions between beads in the MGG.}
\end{figure}

Our theoretical treatment is based on the following fact: as long as the widths of the air channels are much smaller than the channel length, i.e., $w_{1,2}\ll L$, sound wave propagation between junctions can be well described assuming monomode propagation\cite{Depollier,WangPRLNetwork}. Further on, to derive our discrete model we label
the center of each unit cell using the normalized coordinates $m$ and $n$ as shown in Fig.~\ref{fig1}(a). 
Considering the unit cell at position $(m, n)$ and employing
% each junction  is directly connected with its four neighboring juctions through the air channels  $w_1\ll L$ and $w_2\ll L$t waveguides formed 
%
%since the blocks and the two plates are rigid, the air channels can be treated as tubes for sound wave propagation. 
%By considering 
the continuity of flux at each of the four junctions [see supplemental materials (SM)], we derive the following system of discrete equations describing the sound pressure at each junction \begin{subequations} \label{TB}
\begin{eqnarray}
\varepsilon p_\alpha ^{m,n} &=& t p_\beta^{m-1,n}+s p_\beta^{m,n}+t p_\gamma^{m,n+1}+s p_\gamma^{m,n}, \\
\varepsilon p_\beta ^{m,n} &=& s p_\alpha^{m,n} + t p_\alpha^{m+1,n} + t p_\delta^{m,n+1}+s p_\delta^{m,n}, \\
\varepsilon p_\gamma ^{m,n} &=& t p_\delta^{m-1,n}+s p_\delta^{m,n}+s p_\alpha^{m,n}+t p_\alpha^{m,n-1}, \\
\varepsilon p_\delta ^{m,n} &=& s p_\beta^{m,n}+t p_\beta^{m+1,n} +s p_\gamma^{m,n}+ t p_\gamma^{m,n-1}.
\end{eqnarray}
\end{subequations}
In Eqs.~\eqref{TB} $p_i$ with $i=\alpha, \beta, \gamma, \delta$ is the pressure at each junction, 
%the superscripts $m$, $n$ mark the position of each unit cell, 
$t = w_1/(w_1+w_2)$ is the intercellular hopping coefficient, $s=1-t$ is the intracellular coefficient, and the ``energy" term $\varepsilon = 2\cos{2\pi f L/c}$ depends on both the length of each channel $L$ and on the frequency  $f$ while $c$ is the speed of sound in air. To obtain the corresponding dispersion relation we seek solutions in form of Bloch waves as $ p_{i}^{m,n}=p_{i}e^{i k_x m+i k_y n} $, where $k_x$, $k_y$ are the wave vectors along the $x-$ and $y-$directions in the first Brillouin zone (BZ), as shown in the inset of Fig.~\ref{fig2}(a). By substituting the wave solution into Eqs.~\eqref{TB},
we derive the following eigenvalue equation
% a discrete network model for the 2D SSH acoustic network can be obtained,
\begin{equation} \label{eq1}
\begin{bmatrix}
0 & s+t e^{ik_x}  &  s+t e^{-ik_y}  & 0 \\
s+t e^{-ik_x} & 0 & 0 &  s+t e^{-ik_y} \\
s+t e^{ik_y}  & 0 & 0 &  s+t e^{ik_x} \\
 0 & s+t e^{ik_y} &  s+t e^{-ik_x}  & 0
\end{bmatrix} \Psi
=
\varepsilon \Psi,
\end{equation}
where $\Psi=[p_\alpha; p_\beta; p_\gamma; p_\delta] $ is the basis consisting of the pressures of the junctions marked in Fig.~\ref{fig1}(d). %From  Eq.~\eqref{eq1}, it indicates that the sound wave dynamics in a continuous network eventually can be modeled by a discrete tight binding model characterized by the pressures on the junctions of the network. 
It can be seen that, Eq.~\eqref{eq1}) has exactly the same form as the 2D SSH Hamiltonian introduced in Refs.~\cite{WakabayashiPRL,WakabayashiPRB} which confirms that the proposed network is an acoustic realization of the 2D SSH model. Thus, here we directly  bridge the interesting topological properties of the model to the acoustic realm.

At this point we emphasize the fact that the hopping coefficients in Eq.~\eqref{TB} are directly given by the width of the channels in great contrast to TBA where these coefficients are derived as overlapping integrals of wavefunctions. Moreover, another remarkable advantage of our proposed methodology is that a large variety of discrete systems with desired coupling coefficients can be exactly mapped to an acoustic network.
%
%t should be also noticed that, from Eq.~\eqref{eq1}, the propagation of sound waves in a continuous network eventually can be modeled by a discrete network model, which is characterized only by the pressures at the junctions. We believe that the method applied in this work to derive the governing equations can be extended to other acoustic networks, leading to the possibilities of investigation of propagation property of sound waves in more general acoustic systems.

We obtain the dispersion relation by solving the eigenvalue problem of Eq.~\eqref{eq1}.  Note that according to our modeling there is only one  free hopping coefficient $t$ since $s=1-t$. Using the values of  $t=0.714$ and $L=0.125$ m which correspond to the experimental setup, in Figure~\ref{fig2}(a) we show the dispersion curves obtained by solving Eq.~\eqref{eq1}. The band structure is characterized by four propagating branches (gray curves) and two full gaps around $550$ Hz and $900$ Hz. To verify the theoretical results we implement numerical calculations using a finite elements method for a network with  $t=0.714$, $L=0.125$ m and a channel width $w_1/L=0.02$ (not the same as in the experiments) and results are shown in Fig.~\ref{fig2}(b). In this case, by comparing Figs.~\ref{fig2}(a) and (b) we observe that the dispersion relations are almost identical confirming that the wave propagation is very well described by the discrete network model.
Furthermore, we also simulate a network which has the same channel width as our experimental setup i.e. $w_1/L=0.2$ and the corresponding dispersion curve is shown in Fig.~\ref{fig2}(c). In this case, since our main assumption that $w_{1,2}\ll L$ is not well satisfied, the theoretical model and numerical calculation exhibit a shift of between the two dispersion curves. However, in the low frequency regime and around the first band gap that we will focus our study, the discrete network model in Eq.~\eqref{eq1} describes the physical system quite accurately. 

%  same characteristics but 
%regarding $t =0.714$, and $L=0.125$ m, but we investigate two networks %\sout{of} 
%with different width %\sout{of} 
%or the air channels keeping the value  $t =0.714$ unchanged. When $w_1/L=2.5/125$, the dispersion curves are shown in Fig.~\ref{fig2}(b). %\sout{And }
% The experimental setup is designed with channel widths  $w_1=0.025$ m, $w_2=0.01$ m leading to $t=0.714$ while $L=0.125$ m. 
%Fig~\ref{fig2}(c) corresponds to the case when $w_1/L=25/125$. It can be seen that when $w_1/L=2.5/125$, the dispersion curves is almost the same as the one predicted by the network model in Fig.~\ref{fig2}(a), confirming that the propagation of sound waves can be well described by the discrete network model.
%When the width of air channels $w_1$ or $w_2$ are not so small than their length $L$

\begin{figure}[t]
\includegraphics[width=8cm]{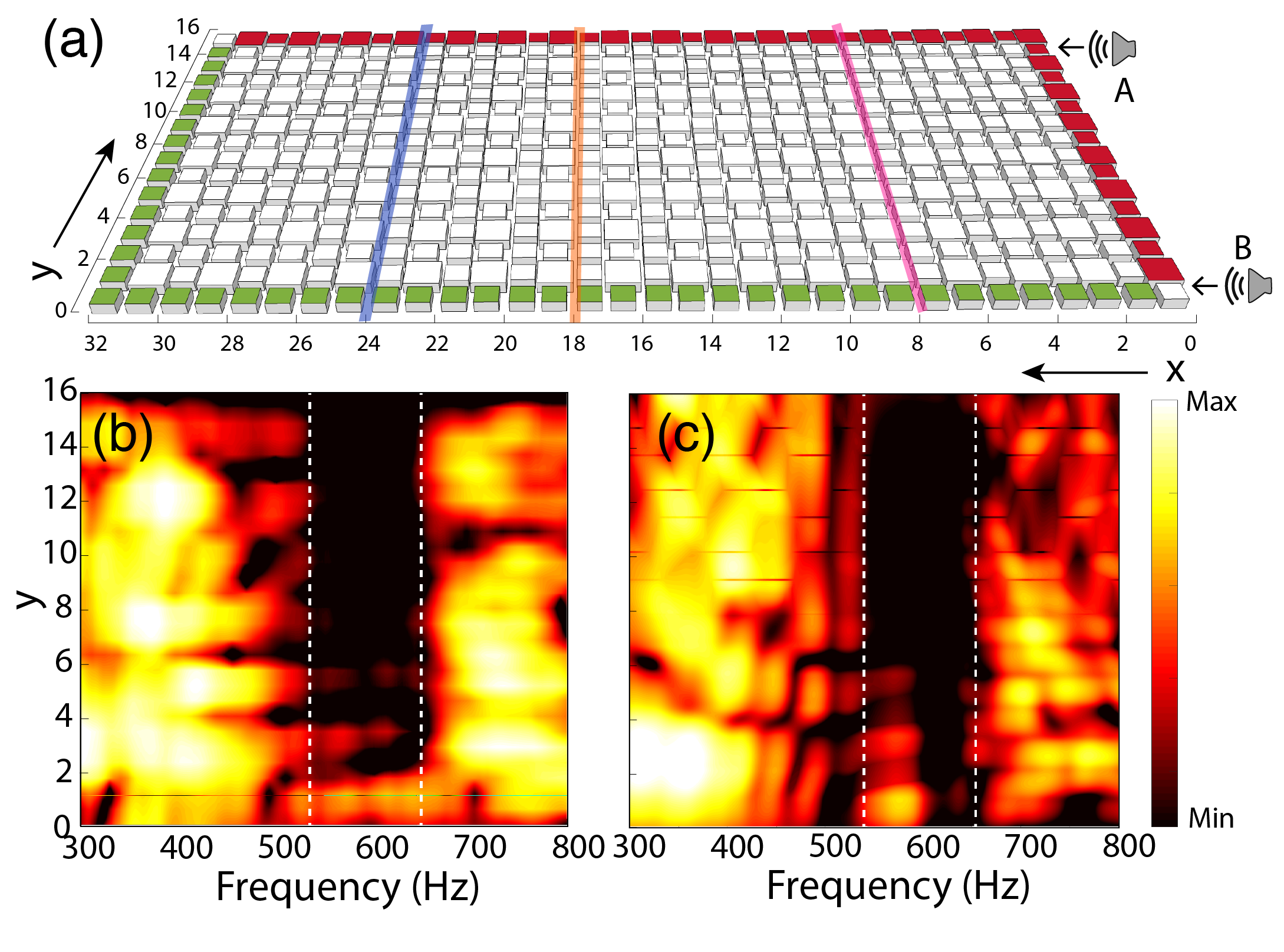}
\caption{\label{fig3} Frequency sweep experiments for the existence of topological band gap. (a) Schematic presentation of experimental setup. The source generated by a loudspeaker can be set at positions $\mathrm{A}$ or $\mathrm{B}$ on the right side of the sample. The free edges in green (red) support (do not support) the the propagation of edge waves. By placing the source at position $\mathrm{A}$, the measurement of band gap is implemented by recording the pressures of line $x=24$ (blue line in (a)). (b) Measured results. (c) Numerical results by simulating the experimental process. }
\end{figure}

One essential property of the 2D SSH model is the appearance of topological edge states when the system acquires a non-trivial topological phase which is achieved by tuning the hopping $t$. In the acoustic network, this transition takes place at the critical case when $w_1=w_2$ ($t=0.5$). When $w_1>w_2$ ($t>0.5$), the network is in the topological non-trivial phase, and can exhibit topological edge waves. Thus our experimental
setup with  $t=0.714$ is designed to fall in the non-trivial phase.
To predict the presence of edge modes theoretically,  we calculate the dispersion relation
of a supercell containing $8$ unit cells with open ends (assuming zero pressure field), see SM. 
The resulting dispersion curves are depicted in Fig.~\ref{fig2}(d).
It can be seen that inside each band gap, there is a degenerate edge wave branch marked with a black line, (see SM).  The dispersion curves  for the supercell obtained by numerical simulations using zero pressure at the boundaries, are shown in Fig.~\ref{fig2}(e), and \ref{fig2}(f) corresponding to the same channel widths as panels (b) and (c) respectively.
Note that from now on we will use zero pressure boundary conditions in all simulations.
Both simulations confirm the appearance of an edge wave branch inside the band gaps
verifying our theoretical prediction.
%As expected, when $w_1/L=2.5/125$, the edge wave dispersion curves also match perfectly to the one of the theoretical model in Fig.~\ref{fig2}(d). When $w_1/L=25/125$, the edge dispersion is shifted slightly compared to the one in Fig.~\ref{fig2}(d). Nevertheless, two edge wave branches still present in the gaps as predicted by the network model. 

\begin{figure*}[t]
\includegraphics[width=13cm]{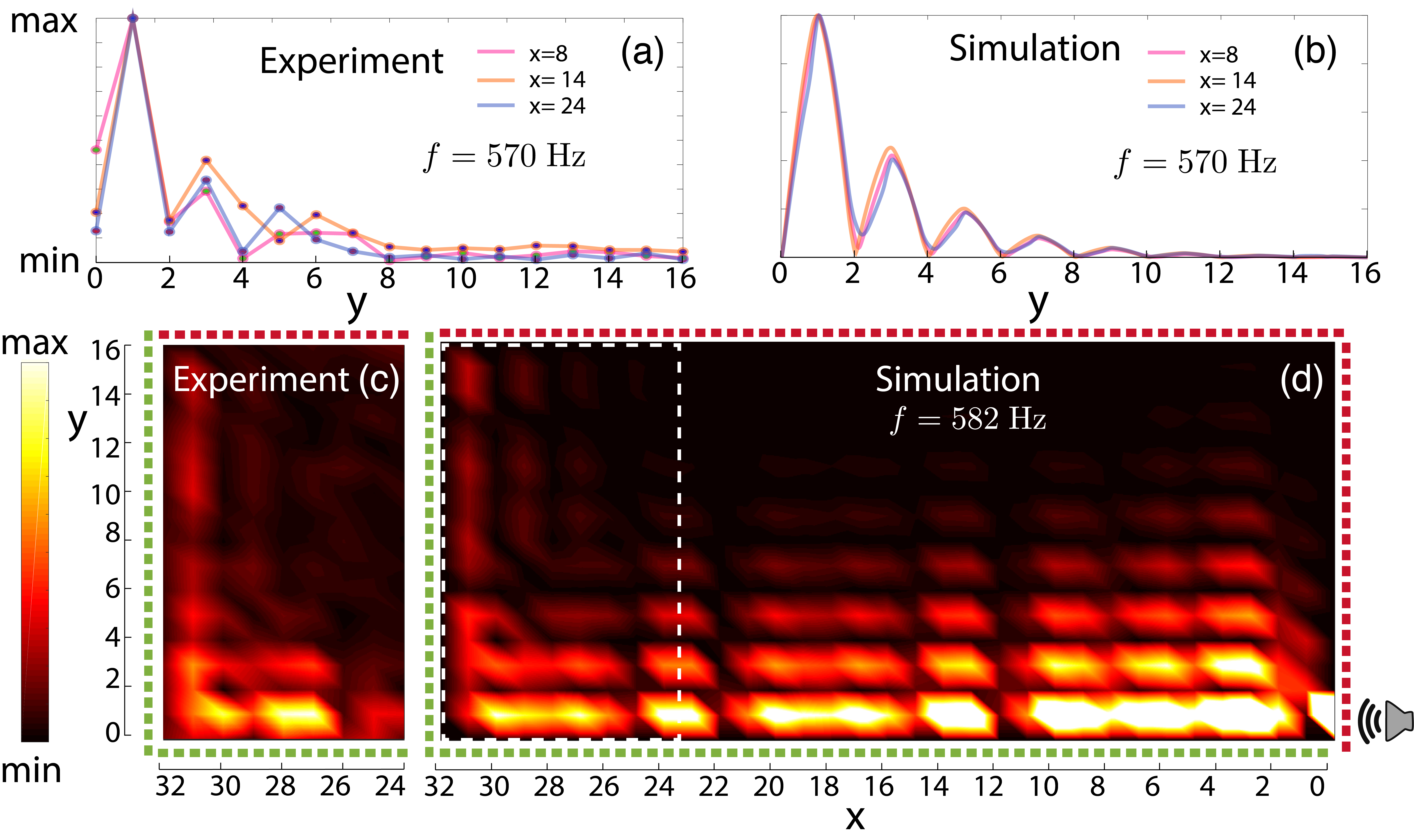}
\caption{\label{fig4} Observation of edge waves.
(a) The edge wave profiles of line 8 (pink), line 14 (orange) and line 24 (blue) when frequency is $570$ Hz. (b) The edge wave profiles obtained by numerical simulation.  
(c) Pressure field distribution obtained by scanning the pressures from line 24 to line 32 in the SSH network when frequency of $582$ Hz is sent to the source. (d) Pressure field distribution obtained by numerical simulation. The white-dashed box corresponds to the same area in (c).} 
\end{figure*}

Let us now  turn to the experimental realization of the SSH network:  its total size is $4$ m $\times$ $2$ m  
and it is constructed {using} building blocks of size $L_1=0.1$ m, and $L_2=0.13$ m, which leads to the air channels of width $w_1=0.025$ m, and $w_2=0.01$ m as shown in Fig.~\ref{fig3}(a).
To emphasise the importance of the edge configuration, 
 two types of edges are simultaneously investigated:
 one supporting edge waves [in green in  Fig.~\ref{fig3}(a)] and the other one with alternative blocks that does not support  edge waves [in red in  Fig.~\ref{fig3}(a)].
%The source 
%We use a loudspeaker  source to generate sound waves from the right side of the network, and we use a microphone to \sout{record the sound wave signals} \textcolor{blue}{to measure the acoustic field inside the network}. Since the size of the microphone (diameter $0.006$ m) is smaller than the sizes of channels $w_1$ and $w_2$, the microphone can penetrate into the network through the channels so that the acoustic pressure \sout{signals} on the junctions can be collected. 
We focus on the first band gap around $\sim 550$ Hz as marked by the orange {area} in Figs.~\ref{fig2}(d)$-$(f).  
We experimentally identify this bandgap and the results are shown in Fig.~\ref{fig3}(b) where the measured pressure amplitude at each junction of line $x=24$ [Fig.~\ref{fig3}(a)] as a function of frequency is plotted. For this experiment, the source is placed at location  $\mathrm{A}$ (see Fig.~\ref{fig3}(a)). Experiments are compared with numerical results shown in Fig.~\ref{fig3}(c), and both confirm the existence of a band gap in the frequency range from $\sim 540$ to $\sim 620$ Hz. A footprint of the edge waves also appears as bright spots inside the bandgap in Figs.~\ref{fig3}(b)-(c) located at $y=0$ (green edge).

% which is embedded on the red edge without edge waves. 
%revealed as bright spots in 
%
%the bulk band gap, depicted with the localisation of pressure field on the green edge.
%

%To experimentally identify this bandgap,  the loudspeaker source is placed at location  $\mathrm{A}$ (see Fig.~\ref{fig3}(a)) which is embedded on the red edge without edge waves. 
%A frequency sweep signal from $300$ Hz to $800$ Hz is generated from the source. 
%We choose line $x=24$ (blue line), and measure the pressures of all the junctions on this line. 
%The measured pressure amplitudes  of all the junctions in line $x=24$ are presented in Fig.~\ref{fig3}(b), and to be compared with
%the numerical results  in Fig.~\ref{fig3}(c). It can be seen that in both cases there is a bulk gap in the frequency range from $\sim 540$ to $\sim 620$ Hz, manifesting the existence of the band gap.
%A footprint of the edge waves also appears in the bulk band gap, with the localisation of pressure field on the green edge.
%Interestingly, in both Figs.~\ref{fig3}(c) and (d), it shows that even in the band gap, {sound pressures can be measured only near the bottom edges (green), suggesting \sout{the} the existence of edge modes on the bottom edges as expected. IT IS NOT CLEAR}

To better characterize the edge waves of the acoustic 2D SSH network we perform additional experiments using a source close
to the green edge [position $\mathrm{B}$ in Fig.~\ref{fig3}(a)]. 
%To inspect the profiles of the edge waves in more details, 
%
%we switch the position of source to place $\mathrm{B}$ as marked in Fig.~\ref{fig3}(a), and a harmonic signal of $570$ Hz is sent from the source. 
%As labelled in Fig.~\ref{fig3}(a), we measure the pressures of junctions on three lines $x=8$ (pink), $x=18$ (orange), and $x=24$ (blue), respectively. 
The normalised edge wave profiles measured at three different lines [$x=8$~(pink), $x=18$~(orange), and $x=24$~(blue) as indicated in Fig.~\ref{fig3}(a)] of the network are presented in Fig.~\ref{fig4}(a). The edge wave, which is excited using a source at  $570$ Hz, 
 is revealed as the acoustic field is localised on the green edge ($y=0$) and is decaying into the bulk. All the experimental profiles are found to be invariant along the $x$-axis and they exhibit the typical pattern provided by SSH models with sub-lattice symmetry \cite{AsbothBook}. Note that in our airborne experiment it is impossible to achieve an exact zero pressure boundary condition due to a slight leakage into free space.  However,  as shown in Fig.~\ref{fig4}(b), exact zero pressure boundary conditions used in numerical simulations lead to the same profiles confirming the robustness of the system with respect to boundary conditions.
%
%The invariance of the profile and the
%
% showing chiral character. \textcolor{red}{In addition, the mode profiles of the three lines exhibit a similar mode pattern, showing the propagation of edge wave at $570$ Hz along the bottom edge.} 
% 
 %The edge profiles obtained  by  numerical simulations are presented in Fig.~\ref{fig4}(b) and are found to be  similar
 %The same results can be obtained from numerical simulations as shown in Fig.~\ref{fig4}(b), where edge wave profiles are clearly presented. 
% 
% 
%To further exploit the properties of the edge waves in the acoustic network, we perform experiments at different frequencies within 
%the bandgap. 
A characteristic field distribution measured at all the junctions located at $x \geq 24$[Fig.~\ref{fig3}(a)] is shown 
in Fig.~\ref{fig4}(c) for a frequency of $582$ Hz. The corresponding numerical results are also shown in Fig.~\ref{fig4}(d), visualising 
the field distribution within the whole network, in good agreement with the experiment.
 We clearly observe that the acoustic field is only localised at the green edges [bottom 
and left of Fig.~\ref{fig4}(c)-(d)] while it vanishes in the other two edges. This is a direct consequence of the particular design of the device which combines two different types of edges: the red edges see the bulk as trivial and the green ones as topological. 

%The ability of the acoustic network to support edge waves on the  These results are a direct manifestation of the topological properties of the SSH model.

%The existence of the edge waves can be seen straightforwardly from the pressure field distribution in the network. In this experiment, we keep the source at position $\mathrm{B}$ but a harmonic signal of $582$ Hz is sent. We scan all the junctions at position $x \geq 24$ (see Fig.~\ref{fig3}(a)) and record the pressures. The measured pressure field distribution is depicted in Fig.~\ref{fig4}(c). And the simulation result of the same experiment is shown in Fig.~\ref{fig4}(d). Both experimental and simulation results clearly show that sound pressures basically locate on the bottom and left edges (green edges), manifesting the propagation of edge wave in the network. 

%In conclusion, in this work we propose an acoustic network to realize the 2D SSH model for sound waves. By applying a discrete model for the network, we analytically predict the existence of topological edge waves on the boundaries. And we experimentally observe the propagation of topological edge waves. The study of this work provides an acoustic demonstration of topological edge waves in the 2D SSH model. It can pave the way for the experimental study of topological properties of sound waves in acoustic systems. 

In conclusion, by applying a 1D approximation in each connection of an acoustic network, we exactly mapped a continuous system to 
the recently proposed 2D SSH model. The latter although with zero curvature is known to support topological edge waves which we observed in this work using an airborne experimental setup. These results show that, despite the fact that the system is open to the free space, the edges are able to support localised waves.
This work provides an acoustic demonstration of topological edge waves in a very simple system that is the 2D SSH model and paves the way for the experimental study of other novel topological phase in 
acoustic systems, such as higher-order topological modes. 

%leakage at the boundary of the system, 
% We analytically predicted the existence of topological edge waves that
%we confirmed by measurements in an airborne experimental setup.
%
%we proposed and designed an acoustic network to realise the 2D SSH model for sound waves. 

\acknowledgments
This work has been funded by the APAMAS, Sine City LMac, and the Acoustic Hub projects.

\end{document}